\documentclass[sigconf]{acmart} 

\AtBeginDocument{%
  \providecommand\BibTeX{{%
    \normalfont B\kern-0.5em{\scshape i\kern-0.25em b}\kern-0.8em\TeX}}}

\copyrightyear{2026}
\acmYear{2026}
\setcopyright{cc}
\setcctype{by}
\acmConference[CHI '26]{Proceedings of the 2026 CHI Conference on Human Factors in Computing Systems}{April 13--17, 2026}{Barcelona, Spain}
\acmBooktitle{Proceedings of the 2026 CHI Conference on Human Factors in Computing Systems (CHI '26), April 13--17, 2026, Barcelona, Spain}
\acmPrice{}
\acmDOI{10.1145/3772318.3790853}
\acmISBN{979-8-4007-2278-3/2026/04}

\usepackage{listings}
\usepackage{subfig}
\usepackage{acmart-taps}
\usepackage{color, soul}
\usepackage[font={normalsize},tableposition=top]{caption}
\usepackage{graphicx}
\usepackage{float}
\floatstyle{plaintop}
\restylefloat{table}
\usepackage{multirow}
\usepackage{colortbl}
\usepackage{tabularx}
\usepackage{lipsum}

\usepackage[T1]{fontenc}
\usepackage[utf8]{inputenc}
\usepackage{CJKutf8}

\usepackage[english]{babel}

\aptLtoX{\newcolumntype{L}[1]{>{\raggedright\let\newline\\ \hspace{0pt}}m{#1}}}{\newcolumntype{L}[1]{>{\raggedright\let\newline\\\arraybackslash\hspace{0pt}}m{#1}}}
\aptLtoX{\newcolumntype{C}[1]{>{\centering\let\newline\\ \hspace{0pt}}m{#1}}}{\newcolumntype{C}[1]{>{\centering\let\newline\\\arraybackslash\hspace{0pt}}m{#1}}}

\newcolumntype{L}[1]{>{\RaggedRight\arraybackslash}p{#1}} 
\newcolumntype{Y}{>{\RaggedRight\arraybackslash}X}

\newcommand{\pnum}{}

\newcommand{\inlinequote}[2]{\textit{``#1''} (#2)}

\usepackage{booktabs}

\newif\ifshowchanges
\showchangesfalse 

\ifshowchanges
    \newcommand\change[1]{\textcolor{blue}{#1}}
    \newcommand\remove[1]{\textcolor{red}{\sout{#1}}}
\else
    \newcommand\change[1]{#1}
    \newcommand\remove[1]{}
\fi

\begin{document}

\title{Toward Pluralizing Reflection in HCI through Daoism}

\author{Aaron Pengyu Zhu}
\affiliation{%
  \department{Division of Industrial Design}
  \institution{National University of Singapore}
  \city{Singapore}
  \country{Singapore}
}
\email{pengyuzhu@u.nus.edu}

\author{Kristina Mah}
\affiliation{%
  \department{Design Lab, School of Architecture, Design and Planning}
  \institution{The University of Sydney}
  \city{Sydney}
  \country{Australia}
}
\email{kristina.mah@sydney.edu.au}

\author{Janghee Cho}
\affiliation{%
  \department{Division of Industrial Design}
  \institution{National University of Singapore}
  \city{Singapore}
  \country{Singapore}
}
\email{jcho@nus.edu.sg}


\renewcommand{\shortauthors}{Zhu et al.}

\begin{abstract}
Reflection is fundamental to how people make sense of everyday life, helping them navigate moments of growth, uncertainty, and change. Yet in HCI, existing frameworks of designing technologies to support reflection remain narrow, emphasizing cognitive, rational problem-solving, and individual self-improvement. We introduce Daoist philosophy as a non-Western lens to broaden this scope and reimagine reflective practices in interactive systems.  Combining insights from Daoist literature with semi-structured interviews with 18 Daoist priests, scholars, and practitioners, we identified three key dimensions of everyday reflection: \emph{Stillness}, \emph{Resonance}, and \emph{Emergence}. These dimensions reveal emergent, embodied, relational, and ethically driven qualities often overlooked in HCI research. We articulate their potential to inform alternative frameworks for interactive systems for reflection, advocating a shift from reflection toward \emph{reflecting-with}, and highlight the potential of Daoism as an epistemological resource for the HCI community.
\end{abstract}



\maketitle

\begin{CJK*}{UTF8}{bsmi}

\section{Introduction} 
In contemporary contexts, where people increasingly encounter the unknown through technological transformation, ecological crises, and public health challenges,  reflection as a meaning-making process becomes more critical than reliance on quick fixes or problem-solution thinking. This echoes Schön's critique of technical rationality~\cite{schon1984reflective} and highlights the importance of reflective engagement for navigating uncertainties of life and sustaining well-being~\cite{mols2016everyday, sengers2005reflective, soden2022modes, light2017design}. Recognizing this, Human Computer Interaction (HCI) has increasingly focused on reflection as a key concern, framing it not only as a conceptual lens but also as a core design goal~\cite{sas2009designing, bentvelzen2022revisiting}. This has led to the development of various interactive systems—including personal informatics (PI) tools~\cite{li2010stage, epstein2015lived, cho2022reflection, choe2017understanding}, mixed reality applications~\cite{wagener2025togetherreflect, ahn2025designing, jiang2022beyond}, AI-driven systems~\cite{kocielnik2018reflection, lyu2025lumadreams}, tangible interfaces~\cite{mols2020everday, rajcic2020mirror, sathya2024attention}, and social media platforms~\cite{bae2014ripening}—typically designed to foster self-awareness, behavior change, or critical perspectives on everyday practices~\cite{li2010stage, fleck2010reflecting, baumer2014reviewing, lim2019facilitating}.

Yet, despite the importance of reflection, the conceptualization of reflection in HCI remains narrow and fragmented~\cite{baumer2015reflective, bentvelzen2022revisiting}. Reflection is frequently defined as a staged process of individual insight or reduced to the act of reviewing data in PI systems~\cite{baumer2014reviewing}. This narrowing of reflection could limit how technologies support people in making sense of experiences and cultivating meaningful practices. Much of HCI research assumes idealized conditions, such as users being motivated, attentive, and cognitively available when engaging in reflective practices~\cite{spiel2018fitter, kersten2017personal}. However, people's everyday situations are often messy, contingent, and saturated with competing demands, leaving little time or cognitive space for deliberate reflection~\cite{dourish2011divining, mols2020everday}. Consequently, the challenge is not merely to prompt reflection, but to support its unfolding within the realities of everyday life. Even when systems successfully elicit reflection, they rarely sustain in-depth reflection and may at times generate unintended negative consequences~\cite{baumer2014reviewing, luo2025reflecting, loerakker2024technology, eikey2021beyond}. Furthermore, existing frameworks are often rooted in Western epistemological traditions~\cite{schon1984reflective, dewey1910how_reprint1997, moon2013reflection, baumer2015reflective, sengers2005reflective}, emphasizing \textbf{\textit{cognition-driven, rational problem-solving, and individual self-improvement}}. This often overlooks the \textbf{\textit{embodied, relational, existential, and culturally situated dimensions}} of reflective practice~\cite{kaptelinin2018technology, nunez2018reflection, ibrahim2024tracking}. 

Research programs, such as critical technical practice~\cite{sengers2005reflective, agre2014toward}, feminist HCI~\cite{bardzell2010feminist}, and entanglement HCI~\cite{frauenberger2019entanglement}, call for alternative ways of knowing and living that resist modernist pursuits of technological solutionism. They emphasize interdependence, openness to uncertainty, situated knowledge, and coexistence with non-human entities~\cite{lu2023participatory, haraway2013cyborg, haraway2013situated, lindtner2018design, livio2022eco, tsing2015mushroom, forlano2017posthumanism}. They challenge the anthropocentirc functionalism that prioritizes efficiency and the appropriation of the world's resources~\cite{bardzell2021wanting}, as well as technological reductionism in dominant Western paradigms~\cite{cunningham2023grounds, van2024secret, frauenberger2016critical, law2015leveraging}. Rather than offering a single cohesive framework, these approaches open up plural alternatives that operate cohesively in theory, values and design practices.

Building on these calls for alternative epistemologies, we turn to Daoism as a non-Western tradition of thought and practice to pluralize how reflection is conceptualized and designed in HCI. Rather than seeking to supplant Western traditions, we approach it as an underexplored perspective that complements and extends ongoing research on reflection. Daoism introduces qualities often overlooked in cognitive or instrumental framings, articulating reflection as a relational process grounded in harmony and ethics~\cite{tan2020revisiting}. Through philosophical concepts such as the ~\emph{Dao} and \emph{Wu-Wei} (effortless action), reflection can be reimagined not as the pursuit of mastery or a solutionist end-point, but as balance, receptivity, and interconnection, opening possibilities for designers to engage plural perspectives~\cite{mah2025rest}. To explore this potential, we examine Daoism's contribution to HCI through the following research question: \emph{how does Daoism conceptualize reflection in ways that move beyond dominant HCI framings?} Building on this analysis, we further discuss how Daoist principles could inspire alternative design directions for technologies that support reflection. 


To ground our inquiry and to explore how Daoism can serve as a method of everyday reflection, we conducted semi-structured interviews with 18 scholars, priests, and practitioners. Our research describes how participants' reflective processes in daily life are supported by Daoist embodied practices and mindsets inspired by its teachings, with a close focus on the body, situation, and perceiving others through becoming permeable. We identified three key components of  reflection: \emph{Stillness}, \emph{Resonance}, and \emph{Emergence}. We present these as critical dimensions of a Daoist reflective paradigm, offering an alternative complement to existing HCI frameworks for reflection-supportive technology design~\cite{fleck2010reflecting, baumer2015reflective, slovak2017reflective}. We discuss design implications inspired by dimensions of Daoist reflective practices and further extend our argument by exploring Daoist epistemology as an alternative foundation for HCI. 


This paper has three main contributions: \textbf{(1) It introduces Daoism as a rich non-Western theoretical perspective that expands and repositions the concept of reflection in HCI to the broader HCI community; (2) It identifies three dimensions of Daoist reflection that could complement existing framework of reflection in HCI;
(3) By drawing on Daoism as an ethico-onto-epistemology~\cite{barad2007meeting}, it articulates design implications for reflective technologies, positioning Daoist reflection as an alternative framework that highlights the entanglement of self, others, and situations.}

\section{Related Work} 

\subsection{What is Reflection?} 

Reflection is often described as a conscious, deliberative cognitive process~\cite{dewey1910how_reprint1997, grant2002self}, yet it can also be improvised and informed by tacit knowledge~\cite{schon1984reflective}. Despite its value across disciplines, the concept remains ambiguous and contested~\cite{clara2015reflection, fleck2010reflecting, thompson2012developing} reflecting its multifaceted nature~\cite{baumer2015reflective}. Moon~\cite{moon2013reflection} notes that activities such as reasoning, reviewing, problem-solving, or critical reflection are often treated as synonymous, with meanings shaped by the goals and theoretical traditions of each field. Consequently, reflection is often invoked with different meanings and for different purposes, depending on the intended goals and the theoretical or disciplinary framework in which it is situated.

In this work, we focus not on the outcomes of reflection but on its notion of reflection as it unfolds in everyday life, rather than in settings intentionally designed to provoke it (e.g., classrooms). We adopt a broad working definition to capture its diverse, situated, and often spontaneous character~\cite{mekler2019framework}. Prior HCI work has addressed everyday reflection as thinking concerning day-to-day activities, with an emphasis on more critical and deliberate thought processes~\cite{mols2016everyday, mols2020everday}. Building on these perspectives, we draw on Donald Schön~\cite{schon1984reflective}, who frames reflection as a spontaneous conversation between individuals and the materials of their practice. His notion highlights the value of reflection in addressing the complexity, uncertainty, instability, uniqueness, and value conflicts that people encounter in everyday life. From this perspective, reflection is not a prescriptive or linear process; rather, it emerges from noticing, reframing, and exploring alternative ways of understanding one's experiences and actions~\cite{cho2021art, tsing2015mushroom, leigh2013reflection}. Accordingly, we adopt the following working definition: \emph{Reflection is a meaning-making process enacted through a conversation between a person and their experiences, an object, others, or a situation.}


\subsection{Tracing Reflection in HCI: Frameworks and Technologies} 
With the third-wave turn in HCI~\cite{bodker2006second}, reflection has gained traction as both an analytical lens and a design goal~\cite{baumer2014reviewing, baumer2015reflective, fleck2010reflecting, sas2009designing}, motivating the development of conceptual frameworks and interactive systems that support reflective practice. 

\subsubsection{Theoretical Frameworks for Designing Reflection}
Reflection in HCI has been articulated through conceptual frameworks that inform design. Fleck and Fitzpatrick synthesized traditions from education and psychology into five levels of reflection~\cite{fleck2010reflecting}: revisiting (R0), reflective description (R1), dialogic reflection (R2), transformative reflection (R3), and critical reflection (R4) by synthesizing traditions of reflective practice from education and psychology into a framework. Drawing philosophical, cognitive, educational, and critical traditions, Baumer proposed ``reflective informatics'', framing reflection through breakdown, inquiry, and transformation~\cite{baumer2015reflective}. Slovák et al.~\cite{epstein2020mapping} extended Schön's reflective practicum into social-emotional learning, offering design guidelines for transformative reflection. These frameworks have become influential in HCI as practical vocabularies for designing technologies, though these guidelines remain challenging to implement~\cite{epstein2020mapping}. 

\subsubsection{Design Space: Interactive Systems for Reflection}
Building on conceptual frameworks, HCI has explored a diverse design space of interactive systems to support reflection. PI tools have long been central, enabling users to collect and visualize data for self-knowledge and behavior change~\cite{li2010stage}. Lived informatics expanded this model by foregrounding the ongoing practices of deciding, tracking, and integrating data into daily life~\cite{epstein2015lived}. Yet the reflection stage itself often remains under-theorized, with systems assuming that insight naturally follows access to data~\cite{baumer2014reviewing}. Later work examined how commercial PI features align with different levels of reflection~\cite{cho2022reflection}, emphasizing subjectivity and customizability, while existential models of behavior change highlight reflection on meaning, context, and lifecycles~\cite{rapp2023exploring}.

In recent years, emerging technologies have expanded the design space of reflection. For instance, XR has been explored as a medium for reflection with its immersive qualities, designed for personal challenges~\cite{wagener2023selvreflect}, emotional expression~\cite{stefanidi2024teen, wagener2025togetherreflect}, mood regulation~\cite{wagener2024moodshaper}, and introspection~\cite{ahn2025designing}. Jiang and Ahmadpour further proposed the RIOR model (Readiness, Immersive estrangement, Observation and re-examination, Repatterning) to conceptualize how VR can scaffold reflective opportunities~\cite{jiang2022beyond}. AI has likewise been investigated for its reflective potential, as dialogue and questioning provide effective scaffolds for reflection. Recent work has leveraged LLMs-based chatbots to support dialogue-driven reflection~\cite{karaturhan2024informing}, explored AI-generated imagery as a means of co-interpreting personal data~\cite{park2025reimagining}, and examined how generative systems can foster self-exploration, self-expression, and collective rituals~\cite{jeon2025letters, shi2024personalizing, rajcic2023message}.

In parallel, slow technology~\cite{hallnas2001slow, odom2012slow} has advanced a different orientation by emphasizing deliberate, unhurried, and embodied engagements with reflection. Artifacts such as Photobox~\cite{odom2012photobox}, the long-living chair~\cite{pschetz2013long}, and Olly~\cite{odom2019investigating} encourage reflection through anticipation, subtle temporality, and re-experiencing everyday materials. Soma design similarly focuses on the ``slow process'' of bodily experience~\cite{tsaknaki2021feeling, loke2018somatic}, helping users to experience and become aware of their bodies in ways intimately intertwined with daily activities. Núñez-Pacheco introduced ``reflection through inner presence'' as a sensitizing concept \cite{nunez2018reflection}, emphasizing subtle bodily sensations and tacit knowledge as resources for more nuanced reflective experiences in soma design and embodied interaction. Synthesizing across such developments, Bentvelzen et al.~\cite{bentvelzen2022revisiting} reviewed prototypes and applications designed to support reflection, identifying four design resources (i.e., temporal perspective, conversation, comparison, and discovery) that illustrate how interactive systems could foster reflection.

\subsubsection{\change{Epistemological Assumptions and Gaps in HCI Approaches to Reflection}}

\change{Although} these diverse systems demonstrate a wide range of strategies and resources for supporting reflection, \remove{yet} they \remove{also} expose critical gaps in how reflection is supported in practice. In a recent meta review of PI systems, Luo et al.~\cite{luo2025reflecting} identified unintended consequences in PI systems, including cognitive burdens, emotional strain, counterproductive behaviors, and tensions with social practices. For instance, by highlighting only isolated information such as consumed calories in numerical form, current weight loss apps have been shown to exacerbate eating disorder behaviors~\cite{eikey2017disorder}. Similarly, Kersten-van Dijk et al. found that standardized visualizations (e.g., the ``10,000 steps a day'' rule) can bias users' interpretations of their data, creating problematic mismatches between system feedback and self-perception~\cite{dijk2016dceptive}. These challenges are further compounded by complex user–technology relationships. Users often perceive technological systems as authoritative \cite{sengers2006staying}, which may lead them to over-trust the perspectives offered by interactive systems for reflection and consequently doubt their own feelings \cite{zhu2025centers}. Additionally, excessive reflection can lead to rumination, trapping users in negative emotional cycles \cite{eikey2021beyond, loerakker2024technology}. 

\change{The negative consequences of these reflective technologies discussed above stem from} \remove{At the level of} \textit{\textbf{underlying design assumptions}}\remove{, interactive systems for reflection often privilege self-centered, goal- designs} \change{that tend to prioritize \textbf{individualistic}, \textbf{self-awareness} and \textbf{goal-oriented approaches}}~\cite{rapp2017know, murnane2018personal, aseniero2020activity}, \change{viewing reflection as a retrospective review of data or as problem-solving aimed at achieving predefined goals.} These approaches neglect sociality, sensitivity to external contexts, and the entanglement of data and user ~\cite{lupton2017feeling, sanches2022diffraction, zhu2025centers, homewood2020removal}, rarely exploring more spontaneous, emergent, and embodied forms of reflection that characterize everyday life~\cite{bhattacharjee2023integrating, rapp2023exploring, nunez2025searching, jiang2022beyond}. 

Fundamentally, reflection in HCI remains shaped largely by Western epistemologies that emphasize analytical, critical, and \remove{solution-oriented approaches} \change{reasoning modes}~\cite{ixer1999theres, edwards2006expertise, sengers2005reflective, ibrahim2024tracking, baumer2015reflective, akama2012way}, \change{typically positioning the mind as the primary site for knowledge and reflection. These models situate reflection within dualistic assumptions of subject–object separation and mind–body dualism~\cite{kant2000critique, dewey1910how_reprint1997}, leaving little room for alternative epistemologies or ontologies, which has resulted in the severe underrepresentation of non-Western historical, cultural, and philosophical traditions in reflection research~\cite{tan2020revisiting, sengers2005reflective} and their consequent marginalization~\cite{akama2020expanding}.} \change{Although slow technology~\cite{hallnas2001slow} and soma design~\cite{hook2018designing} challenge dominant design practices by foregrounding ambiguity, slowness, and embodied experiences, they remain influenced by the epistemological assumptions identified above. Slow technology still frames reflection as a deliberate, mind-centered activity, while soma design, despite challenging mind–body dualism, remains rooted in an individualistic phenomenological stance. In other words, these approaches extend or complicate dominant models of reflection but do not fully challenge the ego- or self-centered cognitive epistemology that positions the reflective self as a bounded, intentional, mind-centered agent.} 
\change{This epistemic boundary clarifies why alternative philosophical traditions are necessary for understanding what reflection means and how technologies might support it.} 

While Buddhist-inspired practices, such as mindfulness and contemplation, have entered the field and been shown to support reflection~\cite{terzimehic2019review, mah2021towards}, these remain largely confined to professional settings~\cite{li2024beyond}. In light of the epistemological challenges and ethical complexities increasingly faced by the HCI research agenda~\cite{frauenberger2019entanglement}, broadening alternative epistemological perspectives \change{to appreciate the scope of reflection in HCI } \remove{on reflection} has become imperative. Reflection as an epistemological commitment~\cite{baumer2015reflective} calls for rethinking how technologies can support it in ways that move beyond existing models. \change{Rather than seeking a ``best'' approach or attempting to synthesize diverse traditions into a unified model, we argue for respecting each knowledge system's cosmological locality while allowing them to coexist, resonate, and occasionally challenge one another.} In this study we extend this agenda by proposing Daoism as a theoretical lens to reconceptualize reflection in HCI and inspire alternative approaches to designing technology for reflection.

\section{Background: Daoism}
Daoist thought, originating in China, centers on the \emph{Dao} (道, path, way) as the guiding principle of existence~\cite{chad2025daoism}. Classical texts such as the \emph{Dao De Jing} (Laozi) and the \emph{Zhuangzi} emphasize alignment with natural rhythms through concepts like \emph{Zi-Ran} (自然, self-soing), \emph{Wu-Wei} (無為, effortless action), and \emph{Yin-Yang} (陰陽, dynamic complementarity)\cite{feng1983history, liu2014dao, serran2013mystical, fang2012yin}. Together, these ideas stress the value of fluidity over rigidity, relational balance over hierarchy, and responsiveness over control. Rather than prescribing fixed rules, Daoism points to an ongoing process of attunement between humans, society, and the natural world. These overarching themes foreground adaptability, interdependence, and transformation, offering generative perspectives for rethinking reflection in HCI. While Western scholarship has often separated Daoism into ``philosophical'' (\emph{daojia}) and ``religious'' (\emph{daojiao}) strands, recent studies emphasize their inseparability, highlighting the unity of Daoist thought and practice~\cite{tan2024reflecting, kirkland1997taoism}.

Our work focuses on the philosophical thoughts related to the concept of reflection embodied in the Laozi and Zhuangzi texts, which are considered the most representative of Daoism~\cite{liu2014dao}. At the same time, we do not strictly exclude their religious or spiritual dimensions, recognizing spirituality as a form of holistic living that can support meaning-making and reflective practice in everyday life~\cite{fook2016finding, botelho2021reflection, markum2024mediating}. In the following, we survey Daoism's influence in contemporary disciplines, introduce its key concepts, and examine how reflection has been understood within this tradition. Our aim is not to provide a doctrinal account of Daoism, but to draw on its central ideas as resources for developing an alternative framework of reflection in HCI, complementing Western perspectives while sensitively introducing non-Western ways of knowing~\cite{hui2019question, akama2012way}.

\subsection{Contemporary Application of Daoism} 
Daoism, as a non-Western intellectual orientation that challenges dominant paradigms, has offered profound insights across diverse disciplinary fields, including wellness~\cite{wang2022yin, yip2004taoism, kohn2005health, ding2020chinese}, education~\cite{hung2013aesthetics, tan2020learning}, and design~\cite{jiang2023exploration, jiang2023implementation, jiang2025designing}. For instance, Jiang and colleagues developed a design tool grounded in Daoist Five Elements to advance more-than-human design approaches move beyond Western-centric discourses.~\cite{jiang2023implementation}; while Hung and Yeh explored how Daoist cultural values could reshape the aesthetics of curriculum in contemporary education~\cite{hung2013aesthetics}. In HCI, prior studies have explored the design of interactive digital experiences inspired by Daoist meditation techniques and Taichi~\cite{Henchoz2021mingshan, Gao2022Meditation}, Mah et al.'s work integrates Daoist concepts of \emph{Wu-Wei} and \emph{Yin-Yang} to reimagine future design agendas for rest in HCI, emphasizing the entanglement between human rest and nonhuman temporalities~\cite{mah2025rest}. However, Daoism remains underexplored as a theoretical or epistemological lens in HCI, where it could offer alternative perspectives for technology design.  

\subsection{Core Concepts: Dao, Zi-Ran, Wu-Wei and Yin-Yang} 

\begin{table*}[htbp]
    \caption{Daoist Key Concepts and Connections with Reflection}
    \centering
    \renewcommand\arraystretch{1.3}{
    \begin{tabular}{m{1.7cm}m{5.3cm}m{7cm}}
    \toprule
\textbf{Concept} & \textbf{Definition} & \textbf{Connection to Reflection} \\
\hline
\emph{Dao} & The origin of the myriad phenomena~\cite{feng1983history}. & Reflection is grounded in embodied alignment with one's environment rather than abstract reasoning. \\
\hline
\emph{Zi-ran} & ``Self-soing'' or spontaneity~\cite{liu2016nature}. & Emphasizes relational existence within reflection. \\
\hline
\emph{Wu-Wei} & Effortless or non-purposive action~\cite{slingerland2007effortless}. & Reflection is not deliberate problem-solving or rational evaluation, but cultivating attunement to circumstances and acting with minimal resistance. \\
\hline
\emph{Yin–Yang} & Interdependence and dynamic complementarity~\cite{fang2012yin}. & Reflection resists fixed conclusions and embraces change and productive tension. \\
    \bottomrule
    \end{tabular}}
    \label{tab:daoistkey}
\end{table*}


\subsubsection{Dao}
\begin{figure*}[!htbp]
    \centering
    \includegraphics[width=0.8\linewidth]{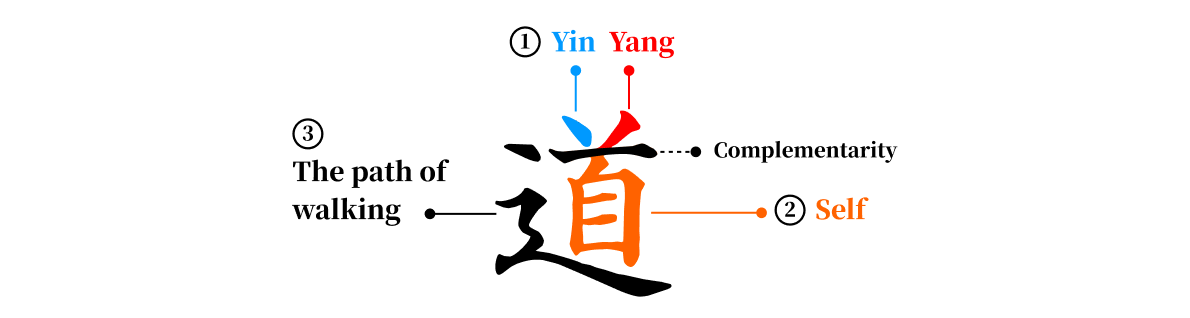}
   \caption{Building on this, turning to the composition of the Chinese character \emph{Dao} (道) can better reveal meanings that go beyond philosophical abstractions. \emph{Dao} (道) consists of two parts. The first part (首) means ``head'', and can also be understood as ``person''. If further broken down (\raisebox{1.1ex}{\scalebox{0.8}[-0.5]{八}}一自), the two dots on the top may be seen as (1) the interdependence of yin and yang, while the horizontal stroke between them represents their integration, manifesting complementarity rather than opposition. (2) The other component is 自, meaning ``self'', which, within the Chinese cosmological view, emphasizes embodiment rather than pure consciousness as in Western philosophy. (3) The other part (\begin{CJK*}{UTF8}{gbsn}辶\end{CJK*}) signifies ``to walk'', or as a noun, ``the path of walking''~\cite{krausova2013beyond}.}

    \label{Dao}
\end{figure*}

As the central concept in Daoism, \emph{Dao} is endowed with metaphysical meaning~\cite{feng1983history}. Scholars have offered diverse interpretations, describing it as ``the origin of the myriad phenomena''~\cite{shih1919outline}. However, many explanations rely on Western philosophy, which fails to fully capture its complex meanings~\cite{liu2014dao}. As expressed at the very beginning of the \emph{Dao De Jing}, \emph{Dao} is of an ineffable nature:
\begin{quote}
\textit{The way that can be spoken of,
Is not the constant way;}
~\cite[Dao De Jing Verse 01]{lau2003daodejing}
 \end{quote}
Building on this ineffability, \emph{Dao} conveys a dynamic system of mutual arising between yin and yang, realized through the unity of the embodied self and its environment (see Fig.~\ref{Dao} for details of the character's composition). \emph{Dao} emphasizes lived experiences, bodily awareness, and attunement to the rhythms of the world, suggesting that reflection lies in alignment with one's environment rather than in abstract reasoning~\cite{krausova2013beyond, merleau-ponty_1945phenomenology}.

This cosmological principle is not only metaphysical but also embodied, taking form as the material vitality known as \emph{Qi} (氣). In Daoist thought, \emph{Qi} is both breath and vital energy, often described as the force that animates the human body and circulates through nature~\cite{graham1989disputers, ye1985outline}. If \emph{Dao} represents the underlying order of the world, then \textit{Qi} is its manifestation in lived experience, sustaining life and connecting all beings~\cite{liu2008body}. 

\subsubsection{Zi-Ran}
Zi-Ran, is often translated as naturalness, yet it differs greatly from the modern concept of nature. The character Zi (自) means ``self'' and Ran (然) means ``so'', so literally it conveys meanings such as ``self-soing'', ``self-going'', ``being free'', or ''spontaneous''~\cite{liu2016nature}, suggesting the spontaneous existence and development of things without artificial interruption or arbitrary control~\cite{liu2014dao, cheng1986environmental}, both the \emph{Dao De Jing} and \emph{Zhuangzi} use metaphor of the infant to express the state of \emph{Zi-Ran}~\cite[Verse 10]{lau2003daodejing}\cite[Miscellaneous Chapter 23]{watson2013complete}. Importantly, the ``self'' in \emph{Zi} is not equivalent to Western notions of individualism. Instead, it transcends a purely human ``state of nature'' and carries deeper ontological significance. As described in the \emph{Dao De Jing}:
\begin{quote}
\textit{Man models himself on earth,
Earth on heaven,
Heaven on the way,
And the way on that which is naturally so.}
~\cite[Dao De Jing Verse 25]{lau2003daodejing}
\end{quote}
This verse shows that \emph{Zi-Ran} is not merely an attribute of nature or individual freedom, but a principle that permeates humans, earth, heaven, and \emph{Dao}. Rather than positioning the self as autonomous or oppositional, \emph{Zi-Ran} emphasizes relational existence, where spontaneity arises through attunement to the larger order of things. 

\subsubsection{Wu-Wei}
\begin{quote}
\textit{The way never acts yet nothing is left undone.
Should lords and princes be able to hold fast to it,
The myriad creatures will be transformed of their own accord.}
~\cite[Dao De Jing Verse37]{lau2003daodejing}
\end{quote}
As Laozi describes in the above text, the non-action of the \emph{Dao} allows all things to transform of their own accord, revealing Wu-Wei as a practical method for attaining the state of \emph{Zi-Ran}~\cite{liu2014dao}. \emph{Wu-Wei} is often translated as ``non-action'', yet its true meaning extends far beyond the literal sense. Slingerland refers to it as ``effortless action'' and “non-purposive action”~\cite{slingerland2007effortless}, while other interpretations include ``never over-doing'', ``no conscious effort'', ``no set purpose'', and ''non-dual action''~\cite{zhu2002wu}. Liu identifies three characteristics of \emph{Wu-Wei}: (1) when things run well, do not interfere; (2) when action is necessary, act without selfish desire;, and (3) always align with the \emph{Dao}, refraining from imposing artificial control~\cite{liu2006introduction}. These traits demonstrate the fluidity and adaptability of non-interfering  action~\cite{liu1996new}, while also carrying an ethical dimension: the spontaneity of one’s actions should not obstruct the natural flow of others. 

Moral subjectivity in Wu-Wei arise not from prescriptive rules but through the dissolution of the self into a state of emptiness and openness~\cite{stokes2016problem, berkson2005conceptions, deprycker2011unself}. The emptiness is not a loss of control, but a release from the demand for a unified, consistent self, embedding the individual in the immediacy of the ever-changing present. In this way, \emph{Wu-Wei} embodies pre-reflective, non-deliberative, and spontaneous~\cite{banner2018wu}. For instance, in martial arts or Tai Chi, movements are not consciously selected but emerge fluidly in response to the opponent from a place of tranquil observation~\cite{knightly2013paradox}. Such an orientation reframes reflection not as deliberate problem solving or rational evaluation, but as cultivating attunement to circumstances and acting with minimal resistance.

\subsubsection{Yin-Yang}
\emph{Yin-Yang} refers to the complementary forces that structure Daoist cosmology~\cite{fang2012yin}. Rather than fixed opposites, \emph{Yin} and \emph{Yang} are interdependent and mutually transforming: heaven, earth, and humanity arise from the continuous interaction~\cite[Verse 42]{lau2003daodejing}~\cite{zhang2008religious}. This dynamic reveals that any state already contains the seed of its opposite~\cite{nisbett2010geography}. 
Laozi described:
\begin{quote}
\textit{Turning back is how the way moves;
Weakness is the means the way employs.}
~\cite[Dao De Jing Verse 40]{lau2003daodejing}
\end{quote}
Here, reversal is not collapse but renewal, and weakness is not absence but generative potential. Such an orientation questions whether reflection is about reaching fixed conclusions, instead framing it as embracing change and tension. 

\section{Methods} 
To complement our conceptual understanding of Daoism from existing literature, we conducted semi-structured interviews with experts (e.g., practitioners, priests, and scholars) who actively engage with Daoism both theoretically and in practice, \change{recent HCI research has similarly engaged philosophers~\cite{wakkary2022two} as well as religious experts and practitioners~\cite{claisse2023keeping, kozubaev2024tuning} through interviews and dialogical inquiry to reveal how philosophical interpretation and spiritual practice can meaningfully inform design and technology.} 

These interviews took place between March and July 2025. Given that foundational Daoist texts such as the \emph{Dao De Jing} and \emph{Zhuangzi} are written in Classical Chinese and are subject to diverse and contested interpretations~\cite{liu2014dao}, we sought to draw upon the living knowledge of experts who work closely with these texts in contemporary contexts. Through these interviews, we aimed to understand how Daoist ideas are interpreted, embodied, and applied in both personal and professional spheres, in order to address our research question. This interpretive approach enabled us to attend not only to established meanings but also to how Daoist philosophies are situated and enacted in lived experience. This research was approved by our institutional review board.

\begin{table*}[htbp]
    \caption{Demographics of the participants}
    \centering
    \label{demographic}
    \renewcommand\arraystretch{1.2}
    \begin{tabular}{@{}clccllc@{}}
    \toprule
    \textbf{ID} & \textbf{Gender} & \textbf{Age} & \textbf{Ethnicity \& \change{Nationality}} & \textbf{\change{Residence}} & \textbf{Daoist Indentity} & \textbf{Years of Practice \/ Study} \\ 
    \midrule
    P01 & M & 40 & Asian \change{(Singaporean)} & \change{Singapore} & Priest & 22 \\
    P02 & M & 25 & Asian \change{(Singaporean)} & \change{Singapore} & Priest & 6 \\
    P03 & M & 26 & Asian \change{(Chinese)} & \change{Australia} & Scholar & 9 \\
    P04 & M & 31 & Asian \change{(Singaporean)} & \change{Singapore} & Priest & 10 \\
    P05 & M & 36 & Asian \change{(Singaporean)} & \change{Singapore} & Priest & 20 \\
    P06 & M & 38 & Asian \change{(Singaporean)} & \change{Singapore} & Priest & 10 \\
    P07 & F & 29 & Asian \change{(Chinese)} & \change{Australia} & Scholar & 3 \\
    P08 & M & 65 & White \change{(Italian)} & \change{Italy} & Priest & 45 \\
    P09 & M & 30 & Asian \change{(Indonesian)} & \change{Indonesia} & Priest & 13 \\
    P10 & M & 39 & Asian \change{(Indonesian)} & \change{Indonesia} & Practitioner & 10 \\
    P11 & F & 42 & White \change{(Dutch)} & \change{Singapore} & Priest & 13 \\
    P12 & M & 63 & Asian \change{(Singaporean)} & \change{Singapore} & Priest & 30 \\
    P13 & M & 71 & Asian \change{(Singaporean)} & \change{Singapore} & Priest & 30+ \\
    P14 & M & 50 & Asian \change{(Singaporean)} & \change{Singapore} & Priest & 10+ \\
    P15 & M & -- & White \change{(Russian)} & \change{Australia} & Practitioner & 10+ \\
    P16 & F & 40 & Asian \change{(Chinese)} & \change{Singapore} & Scholar & 13 \\
    P17 & M & 22 & Asian \change{(Singaporean)} & \change{Singapore} & Scholar/Practitioner & 2 \\
    P18 & F & 35 & White \change{(American)} & \change{Singapore} & Scholar & 13+ \\ 
    \bottomrule
    \end{tabular}
    
    \vspace{0.5em}
    \raggedright
    \footnotesize
    \textit{Note:} M = Male; F = Female; -- = Unknown.
\end{table*}

\change{\subsection{Researcher Positionality}}

\change{Given the cultural and theoretical sensitivity of our work, it is necessary for us to reflect on our positionality. The research team is entirely originally from Asia. The first author as a native Mandarin speaker from China, holds a linguistic and cultural understanding in engaging with Daoist classical texts written in classical Chinese and the philosophical traditions deeply embedded in Chinese culture. His understanding is enriched by situated practice and immersion within Daoist temples in China about half a year. In addition, his broader research examines how nonhuman agencies participate in shaping relationality within marginalized communities through participatory design; this attentiveness to human–nonhuman interdependence further deepens his engagement with Daoist ontological commitments. The second author was born in South East Asia to parents of South East Asian descent.  Her family migrated to Australia when she was a child and she completed her education there. The second author's research background is at the intersection of soma and movement-based research in HCI and culturally informed and embodied epistemologies, inspired by Buddhist and Daoist philosophy. She has been martial arts practitioner of over thirty years, and is a student of both external martial arts, such as karate, and internal Chinese arts, such as \textit{Taijiquan},  \textit{Qigong} and \textit{Neigong}. The last author is originally from East Asia and later lived and received academic training in a Western country, where he developed his grounding in HCI and design research. Although he is not a Daoist practitioner and does not speak Mandarin, he learned classical Chinese characters through public education, and Daoist ideas were part of the broader cultural environment in which he grew up. Through this everyday exposure, he became familiar with philosophical orientations—such as viewing forces in the world as relational and counterbalancing, and understanding personal, social, and environmental realms as interconnected. While these ideas were not part of a formal religious practice, they shaped his intuitive approach to Daoism as a lived philosophy rather than a doctrinal system. His scholarly work is situated in interpretive and critical HCI, particularly reflective design and technologies that support reflective practices. Through this work, he has become aware of limitations in dominant Western approaches to reflection, which motivated the exploration of Daoism as an alternative philosophical framework. 

Therefore, in light of the above research positions, we seek to bridge Daoist philosophy and reflection in HCI with reflexivity, embodiment and epistemic humility~\cite{wakkary2020nomadic}.}

\subsection{Participants \& Recruitment} 
We interviewed 18 experts with diverse forms of engagement with Daoist traditions, including priests, scholars, and practitioners (see Table~\ref{demographic} for participant details). \change{When recruiting participants, we focused on their Daoist identities because these backgrounds shape how individuals understand and engage with Daoist traditions in different ways. Scholars, through systematic study of Daoist classics and philosophical texts, approached Daoism at a conceptual and theoretical level. Practitioners tended to understand Daoism through bodily cultivation, engaging in reflection and realizing the \textit{Dao} through embodied perception. Priests were able to share both philosophical insights and embodied practices, as well as perspectives related to religious rituals (although this paper does not focus on the ritual dimension).} Our definition of ``expert'' was intentionally broad, encompassing individuals with deep experiential or intellectual knowledge of Daoism, gained through academic study, spiritual practice, or applied philosophy. We define Daoist scholars as individuals who have published academic or public-facing work on Daoism or who are affiliated with institutions where they research or teach Daoist philosophy or practice. This group included faculty members and PhD students engaged in relevant research. While no minimum years of experience were required, we prioritized those with sustained engagement through publications, teaching, or formal research. We define Daoist practitioners and priests as individuals who self-identify as practicing Daoists and have engaged in personal or community-based practices, such as rituals, meditation, or involvement with Daoist temples and communities (either online or offline). 

Participants were eligible if they were 21 years or older and fluent in either English or Mandarin. Recruitment was conducted through a combination of purposive and snowball sampling. We identified potential participants by reviewing publicly available information, including academic websites, publications, social media, and online communities where Daoism is discussed or practiced (e.g., Reddit, Telegram group, TikTok). We also contacted local non-governmental organizations related to Daoism and directly reached out to individuals using publicly listed contact details. Snowball sampling was used to extend our outreach to additional eligible participants. Although the study was initiated in Singapore, we did not restrict recruitment geographically, as several participants were based internationally and identified through global networks and online platforms. \change{Because our goal was to understand Daoist perspectives on reflection rather than to compare practices and understandings across national or ethnic groups, we did not treat participants' country of residence or ethnicity as analytical categories. Daoism circulates transnationally and is practiced in diverse cultural contexts, and for this first HCI study engaging Daoist philosophy, we sought to capture a broad range of engagements rather than limit participation to individuals living in specific East Asian cultural environments.}


\subsection{Data Collection: Interviews}

\begin{figure*}[htbp!]
    \centering
    \includegraphics[width=1\linewidth]{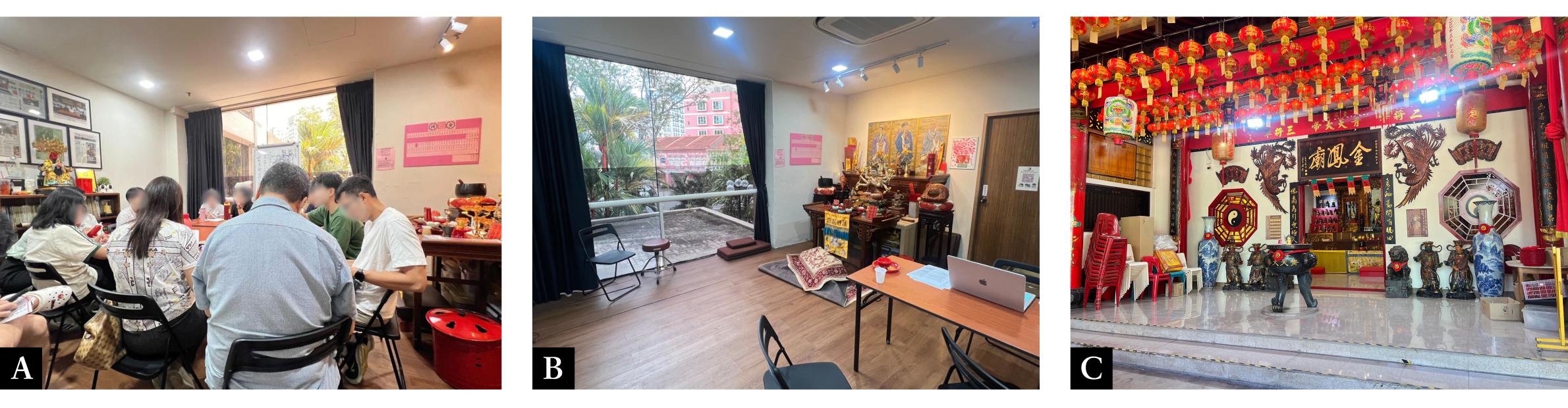}
    \caption{\remove{In person interview settings at the Daoist Association and a local temple in Singapore (The first image shows devotees studying Daoist scriptures.)} \change{Our in person interview settings: \textbf{(A)} Practitioners at the Daoist Association were engaged in a collective scripture study session. We observed their chanting practice and conducted an interview with the priest (P02) leading the recitation after it concluded, this collective chanting practice prompted us to further understand the \textit{dao} as something situated within relationality. \textbf{(B)} The interview setting for another Daoist priest (P01) at the Daoist Association. \textbf{(C)} A Daoist temple in Singapore, where we interviewed three participants (P12-14).}}
    \label{interview}
\end{figure*}

We conducted interviews either in person (see Fig. \ref{interview}) or via Zoom, depending on each participant's preference and availability. Each interview lasted approximately 60 to 90 minutes. All interviews followed a semi-structured format guided by a flexible set of prompts, allowing participants to share their experiences in ways most relevant to their personal and professional engagements with Daoist philosophy. While we explored a common set of themes (e.g., participants' backgrounds, everyday engagement with Daoist principles, everyday reflective practices, ways of sharing these teachings with others, and perspectives on contemporary technologies\footnote{During the interviews, we did not use the specific term ``contemporary technologies''. Instead, we inquired more broadly about their perspectives on using technology in daily life from a Daoist viewpoint and followed up on their responses. Specific categories of technology, such as artificial intelligence, wearable devices, and tracking tools, emerged naturally during these conversations.}), the flow and emphasis of each conversation varied depending on the participant's expertise and lived experience. 

Interviews were conducted primarily in English. However, some participants felt more comfortable discussing Daoist concepts in Mandarin. In such cases, we allowed participants to move fluidly between English and Mandarin during the conversation, depending on what felt most natural for expressing Daoist concepts. At least one member of the research team was fluent in both English and Mandarin, and either one or two team members were present during each interview to ensure accurate communication and effective engagement.

While the concept of reflection was central to our inquiry, we did not impose a rigid or predetermined definition during the interviews. Instead, building on our working definition, we allowed the notion to emerge organically through conversation, following up with open-ended questions to better understand participants' interpretations. This approach reflects our effort to avoid framing reflection solely through dominant Western perspectives, instead foregrounding participants' own vocabularies and worldviews. Accordingly, rather than directly asking about specific Daoist principles, we let related ideas surface naturally; when they did not, we introduced gentle prompts to sustain discussion while remaining attentive to participants' ways of knowing.

With participants' consent, all interviews were audio-recorded and transcribed for analysis. Each participant received a S\$10 (or equivalent) gift card as a token of appreciation. In-person interviews occasionally coincided with community gatherings or ritual ceremonies, offering additional opportunities to observe how Daoist values were enacted in practice. During these encounters, we took field notes and photographs to supplement the interview data and capture insights into the situated and embodied nature of Daoist engagement.

\subsection{Data Analysis}
We conducted a reflexive thematic analysis~\cite{braun2006using} to identify patterns of meaning across the interview data. All interviews were transcribed and, when necessary, translated into English. For transcripts containing mixed Mandarin and English, more than two members of the research team, fluent in Mandarin and English, reviewed the translations to ensure cultural and linguistic accuracy. The initial coding was led by two authors who independently conducted open coding on different subsets of transcripts in ATLAS.ti. Through analytic memoing and regular discussions, the coding process was informed by both inductive insights and sensitizing concepts from relevant literature. The first and third authors then collaboratively reviewed and organized the initial codes into higher-level conceptual groupings. These groupings were iteratively refined through discussion across the team, resulting in the construction of interpretive themes that captured recurring patterns, such as \emph{forgetting} and \emph{listening}. We specifically focused on the different types of practices participants described for elaborating the notion of reflection and their everyday reflective practices, and aligned these with Daoist concepts.

\section{Findings: Dimensions of Reflection in Daoism}

\begin{table*}[htbp]
    \centering
    \caption{Daoist Reflection Components, Methods, Descriptions and Relationship with Existing Framework in HCI}
    \label{tab:daoist-reflection}
    \renewcommand{\arraystretch}{1.4}
    \begin{tabularx}{\textwidth}{>{\raggedright\arraybackslash}p{1.6cm}>{\raggedright\arraybackslash}p{3.4cm}>{\raggedright\arraybackslash}p{4.0cm}>{\raggedright\arraybackslash}X}
    \toprule
    \textbf{Components} & \textbf{Description} & \textbf{Methods} & \textbf{Relationship with Existing Framework} \\[0.3em]
    \midrule
    \multirow{2}[0]{1.6cm}{\textbf{\textit{Stillness}}} & \multirow{2}[0]{3.4cm}{\emph{Stillness grounds reflection as a pre-cognitive, embodied openness that loosens the self and creates space for meaning to appear.}} & \textit{Forgetting}: a reflective practice of releasing attachment to social roles and dualistic distinctions & \multirow{2}[0]{\linewidth}{Existing HCI frames reflection as deliberate reasoning~\cite{fleck2010reflecting}, whereas stillness sees it as openness—less constructing the self, more loosening it.} \\
    \cline{3-3}
    & & \textit{Silence}: an active reflective method that suspends linguistic expression and habitual judgment & \\[0.3em]
    \midrule
    \multirow{2}[0]{1.6cm}{\textbf{\textit{Resonance}}} & \multirow{2}[0]{3.4cm}{\emph{Resonance refers to meaning emerging through inward attentiveness that fosters self-acceptance, together with open and critical awareness of situations.}} & \textit{Listening}: a reflective practice of attending to inner presence, enabling awareness and acceptance of one's present state.& \multirow{2}[0]{\linewidth}{Extending reflection beyond purely cognitive and retrospective models~\cite{fleck2010reflecting, loerakker2024technology} to embodied sensing and experience, with a focus on present-moment experience.} \\
    \cline{3-3}
    & & \textit{Observation}: a reflective practice of maintaining non-judgmental awareness of external situations, balanced with critical vigilance. & \\[0.3em]
    \midrule
    \multirow{2}[0]{2 cm}{\textbf{\textit{Emergence}}} & \multirow{2}[0]{3.4 cm}{\emph{Emergence refers to a dynamic attunement aligning through body, situation and nature rhythms, generating new understanding through adaptation and acceptance of uncertainty.}} & \textit{Attunement}: an active reflective process of being in tune with bodily rhythms, situational flows, and natural cycles. & \multirow{2}[0]{\linewidth}{Emphasizing adaptation and generation rather than control or optimization, while revealing existential and relational dimensions absent in existing reflection frameworks~\cite{fleck2010reflecting}.} \\
    \cline{3-3}
    & & \textit{Becoming}: a reflective practice of perceiving otherness and relationality. & \\[0.3em]
    \bottomrule
    \end{tabularx}
\end{table*}

We identified three interrelated dimensions of Daoist reflection (see Table \ref{tab:daoist-reflection}): \textit{Stillness}, \textit{Resonance}, and \textit{Emergence}. Each dimension expands reflection beyond dominant HCI framings of cognition, \change{individual introspection}, problem-solving, and self-improvement, instead emphasizing emptiness, relationality, and emergence.

Importantly, these three dimensions are all forms of reflection, understood as distinct processes of meaning-making. Each represents a different way of making meaning, without implying a linear progression. From the perspective of \emph{Dao}, where the aim is to live with the \emph{Dao} and cultivate harmony with it, the three dimensions may be interpreted as intertwined stages. These stages are not strictly linear but unfold as part of a trajectory: beginning with \textit{Stillness}, moving toward \emph{Listening} and \emph{Observation} (\emph{Resonance}), and culminating in \emph{Emergence}. 

\subsection{Stillness}

\textbf{\emph{Stillness} is a grounding mode of reflection in which meaning emerges through quieting, emptying, and loosening one's grip on fixed identities and roles (i.e., forgetting).} Rather than discursive problem-solving, Stillness cultivates a receptive awareness that enables reflective practices. 

Participants described \emph{Stillness} as a proactive, embodied practice rather than passive withdrawal. Through meditation, \textit{Qigong}, \textit{Shouyi}, or undivided attention to a single everyday activity, they sought to quiet fluctuations and return awareness to the present. As P06 explained, \inlinequote{your attention is fully on this single action. It is natural. Over time, scattered thoughts calm down. We reduce the burden of our senses and the burden of our mind}{P06}. Several participants framed this as \emph{forgetting} social roles and dualistic distinctions so that a fresh perspective can arise. P02 described: \inlinequote{Maybe our heart suffers because it enters into duality. It is either in the future or the past. We try to bring it back. Now you are not your name or your role. Who are you?}{P02}. For them, \emph{Stillness} sustained attention, stabilized emotion, and clarified judgment without reinforcing a bounded self.

Participants further described silence as an active method that opens reflective space. P03 noted, \inlinequote{Sometimes silence is even more powerful than language. It is a common expression in our intuitive life}{P03}. This resonates with the \emph{Daodejing} on wordless teaching: ``the sage… practices the teaching that uses no words''~\cite[Dao De Jing Verse 2]{lau2003daodejing}. Silence suspended habitual, dualistic evaluation and enabled an integrated perception of situations. Whereas existing HCI frameworks often equate reflection with deliberate reasoning \remove{or retrospective analysis}~\cite{fleck2010reflecting}, Stillness reframes reflection as a state of emptiness and openness \change{rather than}\remove{—less about constructing the self} \change{introspection.}\remove{, more about loosening it.} \change{Deliberate reasoning, in this view, does not deepen reflection but fills the mind with too much thinking, leaving little room for receptive awareness. From this perspective, participants also framed digital technologies as sources of ``noise'' that disrupt reflective space. P01 described mobile technology as a form of ``noise'', sharing: \inlinequote{For me, the phone is more of a distraction—too many nonsense calls all day. So it depends on how people use it. If used properly, it's good. But if you're just scrolling through TikTok, it'll affect you—seeing meaningless things affects your judgment, and your emotions will be different…}{P01}. Rather than supporting openness, such stimulation produces the same restlessness and mental noise that Stillness seeks to quiet. This aligns with longstanding Daoist critiques that technologies can disrupt the natural balance of life. Zhuangzi captures this in his account of the \textit{mechanical heart},\footnote{\change{The term \textit{mechanical heart} corresponds to the Daoist concept of \textit{jixin} (機心), which refers to a mode of reasoning demanded by machines—one that diverts the \emph{Dao} from its pure form and thereby generates anxiety.  In the Zhuangzi, this notion is introduced as follows: ``\textit{Where there are machines, there are bound to be machine worries; where there are machine worries, there are bound to be machine hearts. With a machine heart in your breast, you spoil what is pure and simple; and without the pure and simple, nothing can be truly settled}''.}} a mind overfilled with stimuli rather than emptied for reflection.}

\change{This critique highlights the deep philosophical stakes of Stillness.} Daoist texts frame \emph{Stillness} as cosmologically significant: a return to simplicity and alignment with the \emph{Dao}'s generative void.

\begin{quote}
\textit{There is a thing confusedly formed,
Born before heaven and earth.
Silent and void
It stands alone and does not change,
Goes round and does not weary.}
~\cite[Dao De Jing Verse 25] {lau2003daodejing}
\end{quote}

In this light, \emph{Stillness} is not merely a psychological technique but a way of being situated in the \emph{Dao}'s flow. Classical metaphor such as the infant or the uncarved block illustrate this return to naturalness~\cite[Dao De Jing Verses 15, 28]{lau2003daodejing}. Participant's accounts echoed these sensibilities: reflection began not with discursive review but with cultivated awareness that dissolves ego-bound identities. This form of awareness can resemble early stages of reflection (R0 and R1~\cite{fleck2010reflecting}), yet what participants described was not a substantial self. As P02 summarized, \inlinequote{this awareness is not you; it is just there, observing}{P02}. Thus, \emph{Stillness} reframes reflection as more than self-examination. It is a mode of self-loosening that opens onto the \emph{Dao's} generative grounds, producing responsiveness and relational awareness. 

\subsection{Resonance}

\textbf{\emph{Resonance} is a mode of reflection in which meaning arises through sensitivity to inner states and noticing external situations and others. }It frames meaning-making as relational rather than solitary, widening reflection beyond an individual's cognitive appraisal.

We use the term \emph{Resonance}, drawing on Gan-Ying (感應, ``sympathetic response'') in Chinese cosmology~\cite{hui2019question}. Gan-ying describes how things in the cosmos are mutually responsive, resonating with one another's actions and phenomena. Our findings highlight resonance as a process of meaning-making that unfolds through multidimensional sensory experiences extending beyond the self—particularly through \emph{Listening} (to inner states) and \emph{Observation} (of external situations).

\subsubsection{Listening}

\textbf{\emph{Listening} foregrounds the individual's attention to inner presence, enabling individuals to understand their natural state and to accept any self-state in the present moment.} The concept of listening extends beyond sensory hearing or evaluative judgement~\cite{turner2022qi}. Zhuangzi describes this shift from ear to heart to \emph{qi}:

\begin{quote}
\textit{Unify your attention. Rather than listen with the ear, listen with the heart. Rather than listen with the heart, listen with the energies.} 
~\cite[Inner Chapter 04]{watson2013complete}
\end{quote}

Participants echoed this view by describing attentiveness to breath and bodily sensations. P01 described: \inlinequote{There is the concept of listening, what I call `listening to the breath'. It means not listening to external sounds, but to the voice within oneself.}{P01}. Such \emph{Listening} returned the self to spontaneity and \emph{Zi-Ran} (naturalness)~\cite{matthyssen2025fasting}, resonating with Laozi's image of the supple infant~\cite[Dao De Jing, Verse 10]{lau2003daodejing}. \change{This orientation also illuminated participants' critiques of digital self-tracking. As P17 emphasized, representational data can pull attention away from one's natural state. From her perspective, Daoism seeks technologies that support an authentic sense of \textit{self-soing} and alignment with the present situation, rather than ones that estrange individuals from it: \inlinequote{let's say, screen time and all that, I guess the Daoism would question the, I mean, definitely it can help to a certain extent, but I think the Daoism would think, oh, maybe by having all these screen times and all that, it's already fixating on the number rather than the actual person itself, aligning with the situation at hand.}{P17}. In this sense, listening to inner presence contrasted with the distraction of numerical representations, emphasizing \emph{resonance} to lived experience over fixation on externalized metrics.}

In this state of \emph{Zi-Ran}, listening was also framed as a dialogue with vulnerability. P15 explained: \inlinequote{A certain aspect of the balancing of the psyche occurs simply because of the application of attention to themselves… Within the acceptance, a degree of comfort starts to arise.}{P15}. In this sense, listening enables the practitioner to dialogue with their vulnerability, and through embracing such vulnerability, the individual generates new transformations of the present state of self, being able to \inlinequote{Be at peace with self}{P02}. \emph{Listening} therefore reframes reflection not as deliberate reasoning or cognitive self-assessment, but as an embodied openness that normalizes vulnerability.


\subsubsection{Observation}

\textbf{\emph{Observation} refers to a suspended, non-judgmental awareness of external situations, balanced with critical vigilance.} Zhuangzi illustrates this through the metaphor of the mirror: ``The Perfect Man uses his mind like a mirror—going after nothing, welcoming nothing, responding but not storing''~\cite[Inner Chapter 6]{watson2013complete}. The mirror suggests a mode of perception that reflects circumstances as they are, without clinging to them or carrying forward residues from past encounters. 

Whereas \emph{Listening} turned inward (e.g., attuning to breath, inner states, and vulnerability), \emph{Observation} directed this mirror-like openness outward, toward the surrounding world. Participants described this as maintaining a wider perspective while holding some distance. P06 shared: \inlinequote{I would observe the whole picture of things in a deeper and clearer way. When I am able to see the whole picture, I would decide how to react or how to act…}{P06}.

At the same time, \emph{Observation} was not passive acceptance. It required discernment and vigilance. P07 reflected: \inlinequote{We all have to be vigilant, rather than a kind of blind obedience. What he (Zhuangzi) has been emphasizing… is that he values our ability to make judgments and questions.}{P07}. Thus, observation balanced open reflection, like the mirror's capacity to reflect without storing, with critical questioning that resists blind conformity, while enabling spontaneous responsiveness. In this way, \emph{Observation} reframes reflection not as detached monitoring but as outward attentiveness that resists blind conformity. 


\medskip

Together, \emph{Listening} and \emph{Observation} show how reflection emerges through resonance: inward attentiveness and outward vigilance. Resonance extends reflection beyond individual introspection by grounding it in embodied sensing (\emph{Qi}) and critical openness to circumstances. In contrast to HCI's emphasis on retrospective analysis~\cite{fleck2010reflecting, loerakker2024technology}, resonance emphasizes present-moment awareness as the basis of reflection. 

This orientation aligns with Daoist critiques of rational, instrumental knowledge (\emph{Zhi}, 知)~\cite{lynn2022zhuangzi, wang2025harmonizing, tang2016origins}, and affirms intuitive knowing\footnote{Here we translate it as intuitive knowing, though no single English word fully convey its meaning. The character 明 combines the sun (日) and moon (月), connoting brightness, illumination, and clarity. It is often associated with insight, discernment, or luminous awareness, with an emphasis on embodied and situational knowing. }~(\emph{Ming}, 明). P17 echoed this: \inlinequote{Zhuangzi and Laozi… they criticise Zhi, and they say that Ming is where the reality is… a lot of thoughts will come into my brain and my mind and whatever. I just try to let them be. I try not to label them.}{P17}. Thus, resonance positions reflection not as inferential reasoning~\cite{moon2013reflection, baumer2015reflective}, but as embodied awareness that precedes labeling and analysis, enabling intuitive acceptance of the present moment. In this way, reflection arises through embodied and relational awareness, extending HCI's focus on cognitive or retrospective models toward present-moment experience.

\subsection{Emergence} 

Following \emph{Stillness} and \emph{Resonance}, we conceptualize \emph{Emergence} as a broader mode of reflection that encompasses both \emph{Attunement} and \emph{Becoming}. \emph{Stillness} grounds reflection in quiet receptivity,  \emph{Resonance} widens it through sensory sensitivity, and \textbf{\emph{Emergence} highlights how reflection unfolds as an ongoing process of embodied alignment and relational transformation.} In this sense, reflection arises not as a solitary cognitive act but as an alignment with bodily rhythms, situational flows, and natural cycles, while also dissolving ego-boundaries into interdependent relations with others and the \emph{Dao}.

\subsubsection{Attunement}

\textbf{\emph{Attunement} refers to a mode of reflection in which meaning emerges through alignment.} Participants reflected not only on their inner states but also on their relations with situations and environments, where meaning is created through sensitivity to bodily rhythms, situational flows, and natural cycles.  Whereas \emph{Resonance} described the discovery of meaning in the present moment, \emph{Attunement} highlighted the gradual working with and rebalancing of those meanings over time. 

Participants described aligning body and mind by shifting attention from abstract cognition toward embodied rhythms. P11 explained: \inlinequote{there is not really a limit to what the brain can process, but there is a limit to what the body can process. If you do not align the speed of the brain with the speed of the body, at some point they become disconnected}{P11}. She described the body as a caring medium \inlinequote{to be in line with earth}{P11}, reinforcing its role as a medium for sensing the environment. Laozi similarly highlights this tension of embodiment as both vital and troublesome:  

\begin{quote} 
\emph{The reason I have great trouble is that I have a body.\ When I no longer have a body, what trouble have I?}~\cite[Dao De Jing Verse 13]{lau2003daodejing} 
\end{quote} 

From a Daoist perspective, aligning body and mind constitutes a primary form of individual attunement, grounding reflection as meaning-making.

Participants also emphasized harmonizing with situational flows, exemplified in \emph{Wu-Wei} as attunement with circumstances without deliberate control. P10 described: \inlinequote{It is like following the flow to understand your meaning. Observe the current state of the world, interpret it, then combine that with your strengths and traits to find what you can meaningfully contribute.}{P10}. In this sense, attunement redistributed agency from individual control toward circumstances, allowing responsive spontaneity to emerge. As P05 explained: \inlinequote{Because as long as you don't reject ideas, you have an open mind. You will feel your growth will be really fast, because we don't have our own prejudices, we don't have a strong self-image.}{P05}. Such openness allowed participants to connect broader situational flows with the details of everyday practice~\cite{ameshall2003, whitehead1929process}.

Attunement also required holding instability. Participants frequently adopted a \emph{Yin–Yang} perspective to accept uncertainty, discerning positive meaning within negative experiences. P10, for instance, reflected on a recent financial crisis: \inlinequote{Just this morning I cried with my wife because of financial problems. In the past, I would become very irritated, very sad and disappointed, and start to question why I was so useless or why I had gotten myself into such a difficult situation. But now I tell myself: maybe there is something positive here. Maybe life is trying to teach me a lesson.}{P10} This highlights how attunement involves neither rejecting negative emotions nor being consumed by them, but instead sustaining alignment through balance.

At a broader level, attunement entails synchronizing personal rhythms with those of nature. P06 emphasized that a harmonious lifestyle aligned with natural cycles sustains health, while P16 described cultivating ``nurturing life'' practices in Singapore's climate: \inlinequote{In a place like Singapore where there are no four distinct seasons… As a Chinese, I should sweat in such a hot situation, so I will give myself a chance to sweat a little every day, instead of staying in an air-conditioned room all the time… I want to say that this is a way of life that conforms to my genes.}{P16} Such alignment with natural rhythms illustrates a cosmological orientation in which humans, as products of the Dao, sustain harmony by living in accordance with its generative cycles~\cite{liu2016nature}.

\subsubsection{Becoming}
While attunement emphasizes alignment, Daoist reflection also extends into a broader ontological mode of \emph{Becoming}, grounded in transformation (\emph{Hua}, 化\footnote{the Daoist concept of ``化'' is commonly translated in alignment with the English word ``transformation''.~\cite{tao2011two}}).

\begin{quote} 
\textit{But he didn't know if he was Chuang Chou who had dreamt he was a butterfly,\ or a butterfly dreaming he was Chuang Chou.}~\cite[Inner Chapter 02]{watson2013complete} 
\end{quote}

Zhuangzi's ``Butterfly Dream'' illustrates the meaning of \textbf{\emph{Becoming}: boundaries between self and other collapse, foregrounding transformation not as individual change but as relational becoming.} Unlike HCI's accounts of transformative reflection, often centered on behavioral shifts or new insights~\cite{baumer2015reflective, moon2013reflection, fleck2010reflecting}, Daoist transformation challenges ego-centrism and autonomous subjectivity~\cite{tan2020revisiting, tricker2022cicada}. Zhuangzi's third-person narration dissolves the individual perspective into a relational field where all things co-constitute one another.

Participants expressed this ontological becoming through ethical concern for others. P07 articulated: \inlinequote{Everything and self are actually one. We exist together in the Great One (大化).}{P07}. This reflects Daoism's de-anthropocentric orientation, negating subject–object distinctions and embracing ``self-in-other'' and ``other-in-self'' \cite{graham1989disputers, ho1995selfhood}. P08 echoed this intersubjective stance, describing the presence of an ``invisible Other'' in his martial arts practice: \inlinequote{In my opinion, I think that we cannot think to be alone. Even if we go on a mountain and we want to stay alone, the Other one is always with us. What we call the other one. Maybe it is invisible, but there is.}{P08} is not derived from external moral codes, but arises from an awareness of interdependence and from harmony, rather than conflict, in co-shaping situations with others.

Thus, \emph{Becoming} extends reflection beyond \textit{Attunement}'s alignment of self with rhythms, toward relational transformation with others and the Dao. Participants' reflections revealed how dissolving ego-bound identities enables coexistence with the Other, making Daoist reflective practice inherently ethical and critical. As P08 summarized: \inlinequote{To me, reflection is existence because when I think about what is my reflection in this moment, it is not about myself… the meaning of existence and the connection of existence with something invisible, this to me is the reflection.}{P08}



\section{Discussion}
In this section, we articulate how the Daoist dimensions of reflection---\emph{Stillness}, \emph{Resonance}, \emph{Emergence}---\remove{provide alternative discourses to dominant reflection frameworks in HCI.}\change{extend and reorient prevailing assumptions in HCI approaches to reflection.} We emphasize the significance of \remove{grounding reflective practices as foundation for} \change{emptiness as a generative begining for} meaning-making, and highlight how an attuned mindset may counter negative emotions of rumination while opening possibilities for a turn toward \emph{reflecting-with}. Building on these insights, we outline corresponding design implications. Finally, we point to Daoism, as a non-Western epistemology, as an opportunity to expand the exploratory space of conceptual framings in HCI.

\subsection{Rethinking Reflection: Emptiness, Balance, and Relationality}
\subsubsection{Cultivating Emptiness for Reflection through Stillness} 

Our exploration of Daoist texts and practices indicates that the practices of active forgetting and silence embedded in \emph{Stillness} function as a reflective practice that creates space for meaning-making. Current studies for reflection often generate ready-made insights through AI-driven explanations or visualizations, which constrains reflection to the system's framing of what counts as meaningful~\cite{cho2022reflection}. For example, many PI tools automatically interpret activity logs into labels such as ``productive''~\cite{kim2016timeaware} or ``stressed''~\cite{van2025we}, presenting them as convincing evidence for gaining knowledge. Such representations of experience may constrain opportunities to pause, questions, or reframe what the data might mean in lived contexts. \change{Daoism critiques approaches that amplify human desires and attempt to control or predict outcomes based on intentions and concepts detached from the actual context~\cite{d2025daoist, zhang2025implications}. Within this epistemology, forgetting is considered more vital to one's spiritual life than remembering. It is articulated in far more radical stance: relinquishing the capacity for reasoning, letting go of analytic and cognitive activity, and forgetting accumulated knowledge. Such forgetting embodies a posture of refusal and release—of not being governed by anything~\cite{chen2015memory}.} \remove{In contrast} \change{Correspondingly}, our findings show that Daoist practices of forgetting and silence suspend the impulse to accept ready-made insights. By yielding and emptying, individuals could cultivate space for uncertainty and openness, allowing alternative interpretations of experiences to emerge. 

In Daoism, emptiness holds greater power than fullness, as Laozi describes the \emph{Dao}:
\begin{quote}
\emph{The way is empty, yet use will not drain it.}
~\cite[Dao De Jing Verse 04]{lau2003daodejing}
\end{quote}

The cultivation of \emph{Stillness} aligns with the \textit{Dao}'s ontological movement of reversion~\cite{chen1964does}, a return to a primordial state of non-being and tranquility before the generation of all things. Participants described the forgetting practices of loosening attachment to social identities and seeking a self of \emph{\textbf{Zi-Ran}}, while the functionally approximates Baumer's notion of Inquiry~\cite{baumer2015reflective} or Fleck \& Fitzpatrick's R0 and R1~\cite{fleck2010reflecting}, it differs in method and intent: rather than reviewing past experiences to acquire knowledge, this ``forgetting inquiry'' opens space for the spontaneous emergence of new perspectives\change{, that is, the continual negation of one's assumed identities, knowledge, and positions in order to approach the \textit{Dao}.} Similarly, Akama's articulation of Ma, 間 (emptiness) resonates here, where emptiness fostered sensitivity and relational connection in her work with Indigenous communities by suspending imposition and analysis~\cite{akama2018finger}. \textit{Stillness} likewise serves as a complementary suspension by withholding words and explanation, Fritsch and colleagues show how silence enables deeper attunement to self, others, and environment, supporting awareness, enjoyment, and care~\cite{fritsch2025estrangement}, defining it as filled with sounds, interactions, and possibilities for participation rather than as absence. Thus, the emptiness generated by \emph{Stillness} can be seen as an effective strategy for supporting reflection, even though it often considered as non-reflective practices~\cite{fleck2010reflecting}, as Sartre notes, \textit{``it is the non-reflective consciousness which renders the reflection possible''}~\cite{sartre2022being}, a primordial state that is a foundation that allows one to then step back and reflect from.

\emph{Stillness} constitutes a reflective dimension that enables the spontaneous emergence of meaning, suggesting that reflective technologies might extend beyond scaffolding the reflective phase itself to instead act as flexible enablers that guide individuals into reflective moments. Unlike breakdowns triggered by external systems~\cite{baumer2015reflective}, \emph{Stillness} lies in its proactive orientation—cultivating meaning through the deliberate release of identity and the emptying of thought.

While these findings highlight the methodological significance of \emph{Stillness}, its potential has not yet been fully recognized in designing technologies for reflection. Some efforts in HCI have moved in adjacent directions; for instance, slow technology~\cite{hallnas2001slow} and interactive systems inspired by contemporary mindfulness aim to create a sense of tranquility and guide individuals to focus on the present~\cite{asadi2023calming, prpa2018attending, zhu2017designing, markum2020digital}. To promote individual reflection, such technologies often rely on the materiality of artifacts~\cite{odom2012photobox, asadi2023calming} and specific physical spaces~\cite{jarvela2021augmented}. \change{Meanwhile, some artifacts that support slowness still encourage users to engage in reasoning by introducing deliberately constructed counterfactual patterns that make friction noticeable~\cite{dewey1910how_reprint1997, kant2000critique}—such as SoundMirror and ChatterBox~\cite{hallnas2001slow}. Others evoke reflection by designing temporalities through which one's personal historical data resurfaces in irregular ways~\cite{odom2019investigating, odom2020exploring, odom2012photobox, chen2023exploring}, subtly reinforcing reflection as a mind-centered activity}. \remove{\emph{Stillness} does not deliberately create slow and tranquil moments in life or provide mindfulness-supporting technologies to encourage deliberate reflection. Instead, \textit{Stillness} makes entering reflection possible through active forgetting and linguistic.} \change{Instead, \textit{Stillness} foregrounds \textbf{\textit{Zi-Ran}}, allowing naturalness to emerge rather than arise from deliberate intervention by enabling reflection through intentional forgetting and linguistic silence.} In this sense, we advocate for designing technologies that cultivate \textit{the art of forgetting} \change{for spontaneous reflection}, which aims to fracture externalized information generated by individuals through technology~\cite{ankenbauer2025time, sas2013design}. \change{The assumption in reflective technologies that privileges recording, representing, interpreting, and revisiting information while neglecting forgetting and silence—reveals a computational metaphor of the human mind that frames these necessary, seemingly passive human activities as deficiencies~\cite{bannon2006forgetting, baumer2014reviewing}.} \remove{For example, by regularly clearing tracking data in personal informatics tools, thereby generating retrospective reflection through forgetting, and prompting individuals to become aware of subjective experience review and self-authenticity~\cite{su2016design}} \change{For example, similar to Homewood et al.'s finding that removal creates space for critical reflection~\cite{homewood2020removal}, the \textit{art of forgetting} may advocate designing PI tools that adopt more radical forms of resistance toward data, where individuals' tracked information is automatically cleared at regular intervals after being recorded. This allows users to reflect and interpret their experiences subjectively, thereby moving closer to a sense of self-authenticity~\cite{su2016design}, rather than ``reality'' as interpreted by presets and data representations.} Additionally, shifting reflective prompts from cognitive questions to guiding individuals' bodily sensations, and incorporating estrangement strategies~\cite{wilde2017embodied}, can enable reflection to emerge through defamiliarizing body rather than purely mental activity, \change{while the quieting of external technological noise further moves the individual from outward attention to inward bodily sensing}. 

\subsubsection{Flowing with Uncertainties: Reflective Attunement}

In our findings, Daoist practitioners place importance on \change{attuning with} situation~\cite{ott2024embodiment}. Rather than actively setting and pursuing fixed goals, they adopt a mindset of \emph{\textbf{Wu-Wei}}, flexibly adapting to changes in life, flowing like water, and allowing meaning to emerge naturally in each moment. From this perspective, reflection shifts away from viewing behavioral change as an external goal~\cite{consolvo2009theory, ploderer2014social}, and instead focuses on aligning actions with one's existing rhythms and priorities. This \change{flexible} stance of non-orientation through \emph{Attunement} resembles what Ahmed describes as \emph{disorientation}~\cite{ahmed2020queer}, a willingness to be thrown off course. Also, Daoist practitioners approach this condition through the generative posture of \emph{\textbf{Yin-Yang}}, which enables them to respond more gracefully to life's uncertainty, finding ``joy and excitement in the horror''~\cite{merleau-ponty_1945phenomenology}. In everyday reflection, this position allows them to recognize \change{how negative experiences may tilt toward new forms of growth.}\remove{an inherent potential within negative experiences toward positive transformation.} This Daoist posture invites individuals to dwell with \change{uncertainty} \remove{failure} and attune to its flows, allowing meaning to surface naturally rather than being actively constructed, echoing Laozi in the \emph{Dao De Jing}:

\begin{quote}
\emph{It is on disaster that good fortune perches.}
~\cite[Dao De Jing Verse 58]{lau2003daodejing}
\end{quote}


In HCI, \change{many systems that address negative emotions around reflection are designed to help users manage, regulate, or reframe those emotions toward positive outcomes. Prior work has focused on} \remove{researchers have already explored diverse strategies for negative emotions of reflection, such as} cultivating self-compassion in emotion intervention tools~\cite{smeets2014meeting}, reframing emotional attributes~\cite{eikey2021beyond}, and more recently, Jung et al. proposed a data reinterpretation method~\cite{jung2025thinking}, which enables individuals to contextualize unmet goals in tracking tools as understandable situations rather than personal failures. \change{These are important contributions, yet they often remain aligned with an instrumental logic: negative feelings are treated as obstacles to be resolved on the way to improved functioning, productivity, or well-being. From a Daoist standpoint, this risks narrowing the space of reflection to what can be converted into positive, growth-oriented narratives.} 


\remove{However, from a Daoist perspective, this opens a new pathway for designing technologies that help individuals to cultivate a posture of dwelling with negativity rather than enforcing positive change. This does not imply deliberately designing discomfort~\cite{benford2012uncomfortable}, but rather envisioning reflective technologies that embody more neutral and non-judgmental qualities~\cite{toebosch2024non}, focusing on openness, subjectivity, and fluidity of data representation~\cite{hollis2018being} instead of rigid chart-based formats.} 

\change{Our findings suggest a different orientation: instead of steering users away from negative states, reflective technologies could help individuals dwell-with negativity and uncertainty, attending closely to how negative experiences evolve and change. This attunement-based stance has concrete implications for design. Rather than presenting emotional or behavioral data in rigid, evaluative formats (such as charts that implicitly encode success or failure), systems could offer more open, ambiguous, and non-judgmental representations~\cite{toebosch2024non, hollis2018being}.} For example, designers can explore how image-generating technologies create abstract and ambiguous content that allow users to visualize and explore further allowing users to feel positive tendencies~\cite{park2025reimagining} hidden within feelings of frustration or self-doubt through ambiguity and uncertainty. Importantly, such designs depend on context-sensitive AI strategies~\cite{fang2025mirai, nepal2024mindspace} that capture emotions and behaviors closely tied to users' lived situations and generate representations embedded in everyday experience, rather than simplified numerical indicators~\cite{choe2014understanding, khovanskaya2013everybody, kim2022prediction}, which are more likely to provoke reflection and support dwelling-with. In addition, data physicalization~\cite{jansen2015opportunities, thudt2018self} may provide an effective approach, transforming negative data into perceivable physical forms that connect records with tangible haptic interfaces. For example, by leveraging temperature or deformation to produce sensations of ``comfort''~\cite{daudenroquet2021interoceptive}, technology can foster a sense of acceptance for individuals. 

\change{Furthermore, these orientations invite a rethinking of technological mediation~\cite{verbeek2006materializing}. Rather than functioning as a dominant filter that dictates how the world is revealed or how emotions should be regulated, technology can be designed as a \textit{device for attunement}: something that sensitizes users to the nuances of their situations and supports moment-to-moment responsiveness. This stance challenges the instrumental rationality and the logic of blind action embedded in technologies, as well as their neglect of temporal critique. In this mode, technologies are stripped of their intrinsic value and are seen merely as tools for achieving future goals, leading to to a lack of engagement with what is immediately present~\cite{wenning2011daoism}. In a Daoist \textit{\textbf{Wu-Wei}} oreintation, however, technologies do not aggressively demand attention or enforce explicit lessons. Instead, they recede into the background, offering subtle cues or openings that help individuals stay in touch with the flow of their circumstances. This repositions reflective technologies not as instruments for controlling or correcting experience, but as companions that help people abide in uncertainty, listening, adjusting, and co-creating meaning as situations unfold.}


\subsubsection{As an Assemblage: From Reflection to \textbf{Reflecting-with}}


In HCI, higher levels of reflection, those that involve questioning the self, rethinking the current state of the world, and envisioning alternatives~\cite{baumer2015reflective, fleck2010reflecting}, remain particularly difficult to scaffold through technology. While interactive systems have explored ways to stimulate reflection, most tools focus on instrumental knowledge enhancement, which could help people monitor, manage, or optimize aspects of their lives~\cite{cho2022reflection, luo2025reflecting}, yet they struggle to foster critical reflection~\cite{mezirow1990critical}. This is becoming even more challenging with the growing impulse to delegate critical reflection to AI~\cite{gould2024chattl}, \change{which risks reinforcing cognitive shortcuts rather than cultivating deeper attunement} (consistent with the notions of R3 and R4, which refer to perspective-taking and attention to ethical concerns~\cite{fleck2010reflecting, lee2023we}). 

\remove{In Daoist reflection, the central concern lies in cultivating harmonious situations~\cite{li2025responsive}, emphasizing how the self is situated in interaction with the body, others, and the surrounding context.} 
Daoism positions reflection as an ongoing process of relational emergence and situational harmony ~\cite{li2025responsive}. Our findings reveal that participants loosen and sensitize the self through forgetting and silence (\emph{Stillness}), attend to inner awareness and situational changes (\emph{Resonance}), and strive for attunement with the lively body, context, and nature, thereby refining their awareness of relational \emph{Emergence}. \change{This reflects a Daoist ontology in which the self is never isolated but continually formed through interactions with the world. In this way criticality and ethics arise not from detached analysis but from continuously repositioning one's assumptions and stance} \remove{In this way, criticality, relationality and ethics permeate all stages of Daoist reflection, which does not focus solely on the self but continuously repositions assumptions and stances} through relational engagement, demonstrating a dynamic, embodied, and relational posthumanist subjectivity~\cite{braidotti2013posthuman}. Through such attunement, individuals could achieve a more nuanced integration of the self by interpreting encounters with the world, others, and themselves~\cite{mezirow1990critical}.

Building on this, we propose \emph{reflecting-with} as a sensitising concept \change{and ontological commitment} for technology design. Inspired by Haraway's ``thinking-with''~\cite{haraway2016staying} and more-than-human approaches~\cite{wakkary2021things, oogjes2022weaving, giaccardi2020technology, smith2017designing}, \textit{reflecting-with} emphasizes that reflection unfolds in relation with body, situation, others, and more-than-human entities, \remove{In this view, meaning is not attained through detached retrospection but} emerging through encounters, creating a shared continuity~\cite{zhao2020eastern} and fostering \textit{kinships} with what is in formation~\cite{haraway2016staying}. For instance, the WavData Lamp~\cite{zhong2025investigating} \remove{enables people to interact with an artifact that collects information and music through diverse material forms} \change{predicts individual's music-listening habits through a physical visualization that changes shape and light color, maintaining a close entanglement with users in their everyday environments}, thereby creating an open and generative space for reflection; Areca~\cite{cho2025living}, an IoT-based AI air purifier, records emotional diaries of shared environments, and when people gradually perceive it as a co-dwelling presence rather than mere equipment, such devices foster intimacy and evoke empathy. 


Accordingly, \emph{reflecting-with} suggests that HCI researchers need to move beyond human-centered tools that trigger reflection primarily through data collection and display, and instead consider the agency of technological artifacts in reflection-support systems, attending closely to human–artifact entanglements~\cite{frauenberger2019entanglement, cho2025living, wakkary2022two}. \remove{This involves designing for shared intentionality between humans and artifacts~\cite{somanath2022exploring}, incorporating the environments in which users live, and shaping the indeterminacy of artifacts~\cite{wakkary2016material} to enrich forms of meaning-making.} In this sense, \remove{interactive systems that support reflection should not merely passively collect personal information , but rather transform such information into situational variations through which individuals can be affected,} \change{interactive systems that support reflection should themselves embody the stance of \emph{\textbf{Wu-Wei}} and \emph{\textbf{Zi-Ran}}~\cite{zhang2025implications}, responding fluidly to everyday situations rather than following predetermined principles. Instead of passively collecting personal data and turning it into dashboard, such systems should proactively transform this information into situational variations through which individuals can be affected, constructing \textit{polyphonic narratives} through their material and temporal entanglements with users. This illuminates reflection as an \textit{encounter} rather than a process defined by \textit{who} reflects or \textit{what} is being reflected upon, and is designed to foster shared intentionality between humans and artifacts~\cite{somanath2022exploring} in order to enhance the generativity of meaning-making.} 

Furthermore, another possibility we advocate is the integration of \emph{more-than-human tracking} into PI systems, aligning personal data collection with natural rhythms by incorporating ecological cycles~\cite{chung2025beyong} and \change{multi-dimensional social-contextual information (time factors, seasonal factors, sensory data, spatial factors and values, etc.)~\cite{sun2025traditional}} into the tracking process. \change{Although this approach still presupposes certain data points, its deeper aim is not self-optimization but enabling individuals to cultivate a more nuanced, embodied, and relational reflection between their bodies and environments at a cosmological level. }  

\change{Lastly,} This entangled stance \change{reflected in \textit{reflecting-with}} should not be mistaken for the ``self-disappearance'' often associated with quantified self-paradigms~\cite{rapp2017know}, \change{where individuals can be reduced to mere data points}. Daoist dissolution instead signals an expanded capacity for connection with the broader world, extending reflection into the dimension of care~\cite{de2017matters, wakkary2022two}\change{, when we \textit{reflect-with} rather than reflect on, ensuring an accountability to other entangled entities, human and non-human, fostering responsibility and responsiveness to our shared situations.}


\subsection{Daoist Epistemologies for Rethinking Reflection in HCI}

With regard to reflection, HCI has largely drawn on Western discourses~\cite{dewey1910how_reprint1997, moon2013reflection, schon1984reflective,norman2014things,agre1997computation}. These traditions privilege analytical reasoning and emphasize ``scientific'', ``reductionist'', and ``techno-solutionist'' orientations~\cite{law2015leveraging, lindtner2018design}, often separating knowledge from the lived and embodied contexts of a knower ~\cite{smith2006trying}. This epistemological basis has shaped technologies for reflection that are designed for optimization but struggle to engage users' embodied presence, ethical sensibilities, and aesthetic attunements. \change{In practice, however, reflective moments arise within everyday circumstances that rarely conform to such idealized cognitive models, unfolding instead within fluid, situated, and often non-deliberate modes of experience.}


Recently, HCI scholars have drawn on critical and postmodern perspectives, including feminist Science and Technologies Studies (STS)~\cite{haraway2013cyborg, haraway2016staying}, new materialism~\cite{barad2007meeting, bennett2020vibrant}, and posthumanism~\cite{braidotti2013posthuman, forlano2017posthumanism}, to unsettle binaries such as subject/object, culture/nature, and material/discursive. \change{These theoretical lenses share a commitment to ethics grounded in the immanent, material conditions of the world, recognizing humans as participants in its continual constitution~\cite{geerts2019ethico}. They also signal a shift in HCI toward more entangled, relational, and materially situated modes of inquiry~\cite{dedeouglu2025navigating}.}

Daoism resonates with these critical and postmodern trajectories but also contributes its own distinctive epistemic resources. \remove{particularly its long-standing articulation of a non-dualist ontology and cosmology} \change{Specifically, it offers a \textit{Dao}-centered cosmology in which the \textit{Dao} functions as the generative cosmic force of all things~\cite{miller2003daoism, watts1977tao, liu2014dao, nojonen2019dao}. This cosmology emphasizes a radical entanglement across beings and resisting the ``god-trick''~\cite{haraway2013situated} that abstracts humans from their relations in order to reflect from a distance. By situating all existence within the generative unfolding of the \textit{Dao}, we challenge human exceptionalism and redefines reflection as \textit{Stillness}, \textit{Resonance}, and \textit{Emergence}, through observing, attuning to and co-existing with natural rhythms rather than self-centric controlling. In such a condition, individuals cultivate a cosmological sense of \textit{Gan-Ying}, dissolving boundaries across species and recognize an inseparable condition of self and other. Daoism thus envisions humans as always reflecting and becoming \emph{with} the world~\cite{barad2007meeting}, acknowledging the inherent limits of human agency and the need for humility.}

\change{This Daoist turn also extends accounts of embodied epistemology. While soma design~\cite{hook2018designing} in HCI and Daoism both emphasize a holistic view of body and mind~\cite{shusterman2015somaesthetics}, Daoism places the body within flows of \textit{Qi} and the rhythms of others and nature. Based on this embodied ontology~\cite{freiler2007bridging, slingerland2019mind}, the \textit{shenti} (身体) in Daoism is not an individual's possession, but rather a body that is in \textit{Resonance} with the wider world it is entangled with. It maintains harmony with the rhythms of cosmos~\cite{schroeder2022way, liu2008body, kohn2009being} through inward \textit{Listening} and external \textit{Observation}. As noted by Fritsch et al., soma design currently lacks strategies for engaging with others and the broader ecological system~\cite{fritsch2025estrangement}. A Daoist framing offers such strategies: \emph{Stillness} and \emph{Attunement} cultivate sensitivity not only to one's inner sensation but also to how one's corporeal body is affected by other beings' presence and rhythms~\cite{nunez2024articulating, hook2021unpacking}. This orientation invites inquiry into the enounters that emerge between one's own body and the bodies of others, imagining the body as an emergent, porous, and entangled \textit{cosmological interface}~\cite{homewood2021tracing, peng2025embodying}.} 


\remove{Resonance highlights relational sensitivity to others, situations, and environments. Participants described reflection as listening and sensing through multiple modalities, often dissolving ego-boundaries. This recalls the Daoist story of Woodworker Qing, who forgets his own body to perceive a tree's ``inner heaven''~\cite{chuang2000wandering, ott2024embodiment}. Relational receptivity complements more-than-human HCI's emphasis on ethics, sensitivity, and intimacy in interactions with non-human actors~\cite{nunez2024articulating, nicenboim2025decentering}, inviting designers to create technologies that foreground humility and attentiveness by ``noticing the heaven of materials'' rather than imposing control. } 

\remove{Emergence underscores reflection as an unfolding process rather than a fixed outcome. Participants described adapting fluidly to shifting conditions, findings meaning in aliment with the Dao's rhythms rather than pursuing predetermined goals. This perspective complements cosmotechnics~\cite{hui2019question}, which situates human-tool relations within cosmic harmony. For HCI, this opens up new spaces for the exploration of ``unmaking''~\cite{song2025unmaking, sabie2022unmaking, sabie2023unmaking}\todo{Here, we can discuss on Daoist technological perspective, anti-technology, atnti-efficiency-oriented}, supporting reflection as open-ended transformation rather than linear optimization, fostering intimacy rather than alienation in one's relation to tools and materials, and considering its ``uselessness'' in ethical and aesthetic dimensions\cite{teschner2009technological}. }


\change{Additionally, in current HCI research, technologies for reflection have been designed with individualistic, problem-solving, and goal-oriented aims~\cite{baumer2015reflective, fleck2010reflecting}, privileging epistemologies that are self-centered and instrumental. Such systems often overlook relational and emergent forms of knowing, sidelining the ethical sensibilities that arise through relationality. In contrast, Daoist cosmology~\cite{ames2011confucian, zhao2021dao} is guided by an ethics that respects differences and honors the nature of the ``Other''~\cite{huang2010respecting}. Actions emerge through mutual responsiveness and care, creating a performativity of \textit{poiesis in action} within morally constituted assemblages~\cite{pickering2019poiesis}. In these assemblages, practitioners relinquish self-centered perspectives and preconceived knowledge, adapting with others to the unfolding flow of the \emph{Dao}, a collaborative process of \textit{``way (Dao)-making"}~\cite{zhang2025implications}. This epistemological orientation challenges the individualistic, self-protective stance still implicit in Schön's reflection-in-action~\cite{schon1984reflective}, where the actor maintains a sense of control over objects and situations rather than engaging in a shared agency~\cite{tan2020revisiting}. Daoism instead suggests reflection as an ethico-onto-epistemology, where knowing, doing, and being emerge together through relational regard.}


\change{These commitments also carry practical implications for making, crafting, and Research-through-Design. The Daoist stories of Woodworker Qing, who forgets his own body to perceive a tree's ``inner heaven''~\cite{chuang2000wandering, ott2024embodiment}, and Cook Ding, who follows the natural articulations of an ox by ``\textit{perceiving with his spirit rather than his eyes}''~\cite{chuang2000wandering}, illustrate this vividly. Both artisans embody a humble stance of not-knowing when engaging with materials, sensing and responding to the nature of things rather than imposing one's own intellect as universally valid. Their expertise lies not in control but in adapting fluidly to unfolding situations, entering a process of \emph{becoming-with} the other amidst differences.}

\change{While contemporary more-than-human scholarship highlights material agency, it echoes aspects of this posture~\cite{kim2023surfacing, ofer2021designing, song2022towards, tomico2025constituency}, yet a Daoist orientation suggests an even more delicate integration of one's embodied feelings with non-human rhythms to achieve ``harmony between heaven and humanity" (天人合一). This orientation invites a shift toward the notion of \textit{\textbf{cosmotechnics}} when considering how technologies are designed in HCI~\cite{hui2019question}, a form of technical activity that unifies cosmic order and moral order, treating technology as a type of relation rather than an entity. \textbf{\textit{Cosmotechnics}} foregrounds the ethical grammars embedded in technical action~\cite{engchuan2021lifepatch}, aligning human reflective agency with the self-soing of nature and \emph{Other} beings. This shift moves HCI beyond anthropocentrism toward cosmological attunement and shared becoming.}

\section{Conclusion \& Limitations}

Our study demonstrates how Daoist philosophy can serve as a generative lens for rethinking reflection \change{toward epistemic plurality} in HCI. By examining Daoist concepts alongside participants' accounts, we highlighted alternative dimensions of reflection that move beyond cognitive, deliberative, or outcome-driven framings, and call on HCI researchers to consider embodied, situational, relational, and ethical dimensions when designing interactive systems for reflection. 

We must acknowledge that in this study it was difficult to entirely avoid using Western philosophical terminology when discussing Eastern philosophical thought. However, such comparison or juxtaposition is necessary for readers in the HCI academic community, which is rooted in Western traditions, to understand when introducing non-Western epistemologies, hence, we strive to sensitively use philosophical terms from both epistemologies in our discussion. Furthermore, the Daoist dimensions of reflection we identified may also embody transcendental and embodied qualities that are difficult to articulate textually. Future research could thus envision embodied workshops as a way to move beyond the limitations of textual expression, enabling broader Western audiences lacking cultural background to apprehend these qualities experientially and through alternative forms~\cite{yoo2025translating}. 

In conclusion, Daoism offers underexplored epistemic resources with the potential to open new research agendas in the HCI community. Exploring such Eastern epistemologies is particularly important given that Asia (and other regions of the Global South) has long been positioned as an ``empirical other'' rather than as an ``epistemic other'' capable of challenging the arrogance of universalism~\cite{singh2010connecting}, respecting the cosmological locality of technologies~\cite[p148]{hui2025thinking, guerrero2024cosmovision}. Our study contributes to the recent ontological turn in HCI and STS~\cite{frauenberger2019entanglement} by offering an exploration through Eastern discourses. It acknowledges decentralized ontologies beyond dominant paradigms~\cite{pickering2017ontological}, as a way of ``critically imagining other forms of knowing-being-doing'', and gestures toward a pluriverse of interdependence~\cite{escobar2018designs}.

\bibliographystyle{ACM-Reference-Format}
\bibliography{references.bib}

\appendix

\section{A comparative table of the original Daoist text and English translations}

Since there are numerous English translations of the \emph{Dao De Jing} and the \emph{Zhuangzi}, this paper adopts D.C. Lau's widely circulated translation of the \emph{Dao De Jing}. Table~\ref{tab:dao-verses} presents the selected verses in the order of their appearance in this paper, compared against the Mawangdui manuscript edition (annotated by Wang Bi)~\cite{wangbi2008} as well as the version by Damo Mitchell~\cite{laozi2018daodejing}, which has also been used in the CHI community by Mah et al.~\cite{mah2025rest}. For the Zhuangzi, we likewise chose Burton Watson's English translation ~\cite{watson2013complete} and compare it with the original Chinese text~\cite{chenguying_2007} (see Table~\ref{tab:zhuangzi-verses}).

\begin{table*}[htbp]
\small
\centering
\caption{Comparison of Dao De Jing Verses: Chinese Original and English Translations}
\label{tab:dao-verses}
\renewcommand{\arraystretch}{1.4}
\begin{tabularx}{\textwidth}{>{\raggedright\arraybackslash}p{1.0cm}>{\raggedright\arraybackslash}p{3.8cm}>{\raggedright\arraybackslash}X>{\raggedright\arraybackslash}X}
\toprule
\textbf{Verse} & \textbf{Dao De Jing Chinese Version~\cite{wangbi2008}} & \textbf{D.C. Lau's English Version~\cite{lau2003daodejing}} & \textbf{Damo Mitchell's English Version~\cite{laozi2018daodejing}} \\[0.3em]
\midrule

01 & 
\textit{道可道, 非常道;} & 
\textit{The way that can be spoken of, Is not the constant way;} & 
\textit{The Dao that can be trodden is not an unchanging Dao.} \\[0.5em]

02 & 
\textit{是以聖人處無為之事, 行不言之教}, & 
\textit{Therefore the sage...practises the teaching that uses no words.} & 
\textit{This is why the sage practices Wu-Wei and teaches that which is unspoken.} \\[0.5em]

04 & 
\textit{道冲而用之或不盈。} & 
\textit{The way is empty, yet use will not drain it.} & 
\textit{Dao is void and yet can be drawn upon without ever becoming exhausted.} \\[0.5em]

10 & 
\textit{專氣至柔, 能嬰兒乎?} & 
\textit{In concentrating your breath can you become as supple As a babe?} & 
\textit{Can you gather the Qi until it is as pliable as a newborn infant?} \\[0.5em]

13 & 
\textit{吾所以有大患者, 為吾有身。} & 
\textit{The reason I have great trouble is that I have a body. When I no longer have a body, what trouble have I?} & 
\textit{We only suffer with these things because of our sense of self; if there is no self, how can we suffer so?} \\[0.5em]

15 & 
\textit{敦兮其若樸;} & 
\textit{Thick like the uncarved block;} & 
\textit{They are as simple as an uncarved block of wood.} \\[0.5em]

25 & 
\textit{有物混成,先天地生, 寂兮寥兮,獨立不改, 周行而不殆, 可以為天下母. ... 人法地, 地法天, 天法道, 道法自然。} & 
\textit{There is a thing confusedly formed, Born before heaven and earth. Silent and void It stands alone and does not change, Goes round and does not weary. ... Man models himself on earth, Earth on heaven, Heaven on the way, And the way on that which is naturally so.} & 
\textit{Before Heaven and Earth was chaotic emptiness; it exists still, since it is everlasting. It is both formless and soundless. It does not grow weary and operates endlessly. It is the mother of Heaven and all beneath it. ... But all men must follow the way of Earth. Earth must follow the way of Heaven, Heaven must follow the way of Dao. In turn, Dao must follow the way of true self-nature.} \\[0.5em]

28 & 
\textit{常德不離, 復歸於嬰兒。} & 
\textit{Then the constant virtue will not desert you And you will again return to being a babe.} & 
\textit{This ravine beneath Heaven brings forth De. This is the state of reverting to infancy.} \\[0.5em]

37 & 
\textit{道常無為而無不為, 侯王若能守之, 萬物將自化。} & 
\textit{The way never acts yet nothing is left undone. Should lords and princes be able to hold fast to it, The myriad creatures will be transformed of their own accord.} & 
\textit{Dao adheres to Wu-Wei, and then all things are done. If the princes and kings could only adhere to Dao, the myriad beings would be spontaneously transformed.} \\[0.5em]

40 & 
\textit{返者, 道之動; 弱者, 道之用。} & 
\textit{Turning back is how the way moves; Weakness is the means the way employs.} & 
\textit{'Returning' is the motion of Dao. Dao works through the receptive.} \\[0.5em]

42 & 
\textit{道生一, 一生二, 二生三, 三生萬物。} & 
\textit{The way begets one; one begets two; two begets three; three begets the myriad creatures.} & 
\textit{From Dao comes the one, from the one comes two. Two produces three, and from the three emerge the myriad beings.} \\[0.5em]

58 & 
\textit{禍兮福之所倚。} & 
\textit{It is on disaster that good fortune perches.} & 
\textit{Good fortune is rooted in calamity.} \\[0.5em]

\bottomrule
\end{tabularx}
\end{table*}

\begin{table*}[htbp]
\small
\centering
\caption{Comparison of Zhuangzi Passages: Chinese Original and English Translation}
\label{tab:zhuangzi-verses}
\renewcommand{\arraystretch}{1.4}
\begin{tabularx}{\textwidth}{
  >{\raggedright\arraybackslash}p{1.2cm}
  >{\raggedright\arraybackslash}X
  >{\raggedright\arraybackslash}X}
\toprule
\textbf{Chapter} & \textbf{Zhuangzi's Chinese Version~\cite{chenguying_2007}} & \textbf{Burton Watson's English Version~\cite{watson2013complete}} \\
\midrule

02 & 不知周之夢為胡蝶與，胡蝶之夢為周與？ & \textit{But he didn't know if he was Chuang Chou who had dreamt he was a butterfly, or a butterfly dreaming he was Chuang Chou.} \\[0.4em]

04 & \textit{若一志, 無聽之以耳而聽之以心，無聽之以心而聽之以氣！} & \textit{Unify your attention. Rather than listen with the ear, listen with the heart. Rather than listen with the heart, listen with the energies.} \\[0.4em]

06 & 至人之用心若鏡，不將不迎，應而不藏，故能勝物而不傷。 & \textit{The Perfect Man uses his mind like a mirror—going after nothing, welcoming nothing, responding but not storing.} \\[0.4em]

12 & 有機械者必有機事，有機事者必有機心。機心存於胸中，則純白不備。& \textit{Where there are machines, there are bound to be machine worries; where there are machine worries, there are bound to be machine hearts. With a machine heart in your breast, you've spoiled what was pure and simple; and without the pure and simple.} \\[0.4em]

23 & \textit{「能兒子乎？」兒子動不知所為，行不知所之，身若槁木之枝而心若死灰。} & \textit{``Can you be a baby?'' The baby acts without knowing what it is doing, moves without knowing where it is going. Its body is like the limb of a withered tree, its mind like dead ashes.} \\

\bottomrule
\end{tabularx}
\end{table*}

\end{CJK*}
\end{document}